\documentstyle[preprint,eqsecnum,aps,epsf]{revtex}	

\newif\iftightenlines\tightenlinesfalse
\tightenlines\tightenlinestrue

\begin{document}
%
\def\mol{Mol}
\def\etmiss{E\llap/_T}
\def\eslt{E\llap/_T}
\def\esl{E\llap/}
\def\msl{m\llap/}
\def\to{\rightarrow}
\def\te{\tilde e}
\def\tmu{\tilde\mu}
\def\ttau{\tilde\tau}
\def\tl{\tilde\ell}
\def\ttau{\tilde \tau}
\def\tg{\tilde g}
\def\tnu{\tilde\nu}
\def\tell{\tilde\ell}
\def\tq{\tilde q}
\def\tu{\tilde u}
\def\tc{\tilde c}
\def\ts{\tilde s}
\def\tb{\tilde b}
\def\tst{\tilde t}
\def\tt{\tilde t}
\def\tw{\widetilde W}
\def\tz{\widetilde Z}

\hyphenation{mssm}
%
\preprint{\vbox{\baselineskip=14pt%
   \rightline{FSU-HEP-971104}\break 
   \rightline{UH-511-887-97}\break
   \rightline{UM-TH-97-19}\break
}}
%
\title{$b\to s\gamma$ CONSTRAINTS ON THE 
MINIMAL SUPERGRAVITY MODEL WITH LARGE $\tan\beta$}
\author{Howard Baer$^1$, Michal Brhlik$^2$, Diego Casta\~no$^3$
and Xerxes Tata$^4$}
\address{
$^1$Department of Physics,
Florida State University,
Tallahassee, FL 32306 USA
}
\address{
$^2$Randall Laboratory of Physics,
University of Michigan,
Ann Arbor, MI 48109 USA
}
\address{
$^3$Saint Leo College,
St. Leo, FL 33574 USA
}
\address{
$^4$Department of Physics and Astronomy,
University of Hawaii,
Honolulu, HI 96822, USA
}
\date{\today}
\maketitle
\begin{abstract}

In the minimal supergravity model (mSUGRA), as the parameter $\tan\beta$
increases, the charged Higgs boson and light bottom squark masses decrease, 
which can potentially increase contributions from
$tH^\pm$, $\tg\tb_j$ and $\tz_i\tb_j$ loops in the decay $b\to s\gamma$. 
We update a previous QCD improved $b\to s\gamma$ decay calculation to
include in addition the effects of gluino and neutralino loops. 
We find that in the mSUGRA model, loops involving charginos also increase, 
and dominate over $tW$, $tH^\pm$, $\tg\tq$ and $\tz_i\tq$ 
contributions for $\tan\beta\agt 5-10$.
We find for large values of $\tan\beta \sim 35$ that most of the parameter space 
of the mSUGRA model for
$\mu <0$ is ruled out due to too large a value of branching ratio 
$B(b\to s\gamma )$. 
For $\mu >0$ and large $\tan\beta$,
most of parameter space is allowed, although the regions with the least 
fine-tuning (low $m_0$ and $m_{1/2}$) are ruled out due to too {\it low}
a value of $B(b\to s\gamma )$. 
We compare the constraints from $b\to s\gamma$ to constraints from the 
neutralino relic density, and to expectations for sparticle discovery at
LEP2 and the Fermilab Tevatron $p\bar p$ colliders.
Finally, we show that
non-universal GUT scale soft breaking squark mass terms
can enhance gluino loop contributions to $b\to s\gamma$ 
decay rate even if these are diagonal.

\end{abstract}

\medskip



\section{Introduction}

Models of particle physics including weak scale supersymmetry (SUSY) are
amongst the most promising candidates\cite{haber} 
for new physics at the TeV scale.
Of this class of models, the minimal supergravity (mSUGRA) model stands out as 
providing one of the most economic explanations for the diversity of soft 
supersymmetry breaking terms in the SUSY Lagrangian\cite{drees}. 
In this model\cite{sugra}, 
supersymmetry is communicated from a hidden sector (whose dynamics
leads to the breaking of supersymmetry) to the observable sector 
(consisting of the fields of
the Minimal Supersymmetric Standard Model or MSSM) via gravitational
interactions. With the assumption of canonical kinetic terms for
scalars in the Lagrangian, this leads to a universal mass $m_0$ for all
scalar particles at some high scale $Q$, usually taken to be $M_{GUT}$. 
At $M_{GUT}$, gaugino 
masses and trilinear terms are assumed to unify at $m_{1/2}$ and $A_0$,
respectively. These parameters, along with the bilinear soft term $B$,
provide boundary conditions for the renormalization group
evolution of the various soft terms from $M_{GUT}$ to $M_{weak}$.
Requiring in addition radiative electroweak symmetry breaking leaves a
rather small parameter set
\begin{equation}
m_0,\ m_{1/2},\ A_0,\ \tan\beta\ {\rm and}\ sign(\mu ),
\end{equation}
from which the entire SUSY particle mass spectrum and mixing parameters may
be derived.

The flavor changing neutral current decay of the bottom quark $b\to s\gamma$
is well known to be particularly sensitive to new physics effects. New 
weak scale particles ({\it e.g.}, a chargino $\tw_i$ and top squark $\tst_j$)
could give loop contributions which would be comparable to the Standard Model
(SM) $tW^-$ loop amplitude. Measurements from the CLEO experiment\cite{cleo}
restrict the inclusive $B\to X_s\gamma$ branching ratio to be 
$B(B\to X_s\gamma )=(2.32\pm 0.57\pm 0.35)\times 10^{-4}$, where
$1\times 10^{-4}< B(B\to X_s\gamma )<4.2\times 10^{-4}$ at 95\% CL.
Many analyses have been performed\cite{masiero,previous,bsg1} 
which compare theoretical
predictions of SUSY models to experimental results.

In a previous report\cite{bsg1}, predictions of the $b\to s\gamma$ decay
rate were made as functions of the mSUGRA model parameter space. 
In this study, a number of QCD improvements were incorporated into the
calculation which reduced the inherent uncertainty of the $b\to s\gamma$
decay rate predictions due to
the QCD scale choice from $\sim 25\%$ down to $\sim 9\%$. SUSY contributions
to the $b\to s\gamma$ decay amplitude included $tW$, $tH^+$ and $\tw_i\tq_j$
loops. Results were presented for $\tan\beta =2$ and $10$, and for both
signs of $\mu$. For $\mu <0$, large regions of parameter space were 
excluded, especially for $\tan\beta =10$. For $\mu >0$, all the parameter
space scanned was allowed by CLEO data: in fact, for some ranges of
parameters, the model predicts values of $B(b \to s\gamma)$ close to
the central value measured by the CLEO Collaboration.

Recently, sparticle mass spectra
and sparticle decay branching ratios in the mSUGRA model 
have been reanalysed for large values of
the parameter $\tan\beta$\cite{bcdpt}. 
In the mSUGRA model, the range of $\tan\beta$ 
is typically $1.6\alt \tan\beta \alt 45-50$, where the lower limit
depends somewhat on the precise value of $m_t$. 
For $\tan\beta\agt 5-10$, $b$ and $\tau$ Yukawa couplings become non-negligible
and can affect the sparticle mass spectrum and decay branching fractions.
The upper and lower limits on $\tan\beta$ are 
set by a combination of requiring a valid solution to radiative electroweak
symmetry breaking, and requiring perturbativity of third generation Yukawa
couplings between the scales $M_{weak}$ and $M_{GUT}$. Some optimization 
of scale choice at which the one-loop effective potential is minimized
was found to be needed in Ref. \cite{bcdpt} in order to gain stable
sparticle and Higgs boson mass contributions. This scale optimization 
effectively includes some portion of two-loop corrections
to the effective potential\cite{carena}.
%
It was shown that the mass of
the pseudoscalar Higgs boson $A^0$, and the related masses of 
$H^0$ and $H^+$, suffer a sharp decrease as $\tan\beta$ increases.
In addition, the masses of the lighter tau slepton $\ttau_1$ and 
bottom squark $\tb_1$ also decrease, although less radically.
Naively, one might expect corresponding increases in the loop
contributions to $b\to s\gamma$ decay involving $\tb_1$ and $H^+$.

Indeed, Borzumati has shown in Ref. \cite{francesca} that as 
$m_{H^+}$ decreases, the charged Higgs 
contribution to $b\to s\gamma$ decay does increase. However, for  
large values of $\tan\beta$, the chargino loop contributions increase even more
dramatically, and dominate the decay amplitude. 
She further notes that at intermediate to large $\tan\beta$ values,
there is a non-negligible contribution from $\tg\tq$ loops.

In this paper, we re-examine constraints on the mSUGRA model from
$b\to s\gamma$ decay at large $\tan\beta$. In doing so, we incorporate 
several improvements over previous analyses.
\begin{itemize}

\item We present our analysis using updated mSUGRA mass predictions for
large $\tan\beta$, using a renormalization group improved one-loop
effective potential with optimized scale choice 
$Q=\sqrt{m_{\tst_L}m_{\tst_R}}$. We use an updated value of top mass
$m_t=175$ GeV.

\item We include in this analysis contributions from $\tg\tq_j$ and
$\tz_i\tq_j$ loops. These contributions require knowledge of the full
squark mixing matrices, and hence an improved calculation of 
renormalization group evolution of soft SUSY breaking parameters.

\item As in Ref. \cite{bsg1}, we include the dominant 
next-to-leading order (NLO)
virtual and bremsstrahlung corrections to the operators mediating
$b\to s\gamma$ decay at scale $Q\sim m_b$. 
In addition, we include NLO RG evolution of
Wilson coefficients between scales $M_W$ and $m_b$. We also include
appropriate renormalization group evolution of Wilson coefficients
at high scales $Q>M_W$ for $tW$, $tH^+$ and $\tw_i\tq_j$ loops
following the procedure of Anlauf\cite{anlauf}. The corresponding
RG evolution of Wilson coefficients for $\tg\tq_j$ and $\tz_i\tq_j$
loops is not yet available.

\item We compare our results to recent calculations at large $\tan\beta$ 
of the neutralino
relic density and direct dark matter detection rates for the mSUGRA model.
\end{itemize}

In Sec.~II of this paper, we present some details of our
calculations, especially those regarding the inclusion of
$\tg\tq_j$ and $\tz_i\tq_j$ loops. In Sec.~III, we present QCD improved 
results for the $b\to s\gamma$ branching fraction in mSUGRA 
parameter space for large $\tan\beta$. In this section, we also make 
comparisons with cosmological and collider search expectations.
In Sec.~IV, we relax some of the assumptions
of the mSUGRA framework to see whether $\tg\tq_j$ loops can
become large or even dominant. This question is important when considering
the model dependence of our results. 

\section{Calculational details}

The calculation of the width for $b\to s\gamma$ decay proceeds by
calculating the loop interaction for $b\to s\gamma$ within a given model
framework, {\it e.g.}, mSUGRA, 
at some high mass scale $Q\sim M_W$, and then matching
to an effective theory Hamiltonian given by
\begin{equation}
H_{eff}=-{4G_F\over \sqrt{2}} V_{tb}V^*_{ts}\sum_{i=1}^8 C_i(Q )O_i(Q ),
\label{eq1}
\end{equation}
where the $C_i(Q )$ are Wilson coefficients evaluated at scale $Q$,
and the $O_i$ are a complete set of operators relevant for the process
$b \to s\gamma$, given, for example,
in Ref. \cite{gsw}.
All orders approximate QCD corrections are included via renormalization group
resummation of leading logs (LL) which arise
due to a disparity
between the scale at which new physics enters the $b\rightarrow s\gamma$ loop
corrections (usually taken to be $Q\sim M_W$), and the scale at which
the $b\rightarrow s\gamma$ decay rate is evaluated ($Q\sim m_b$).
Resummation then occurs when we solve the renormalization
group equations (RGE's) for the Wilson coefficients
\begin{equation}
Q {d\over dQ} C_i(Q )=\gamma_{ji} C_j(Q ),
\label{eq2}
\end{equation}
where $\gamma$ is the $8\times 8$ anomalous dimension matrix (ADM),
and
\begin{equation}
\gamma={\alpha_s\over 4\pi}\gamma^{(0)}+({\alpha_s\over 4\pi})^2
\gamma^{(1)}+\ldots .
\label{eq3}
\end{equation}
The matrix elements of the operators
$O_i$ are finally calculated at a scale $Q\sim m_b$ and multiplied by
the appropriately evolved Wilson coefficients to gain the final
decay amplitude.
The dominant uncertainty in this leading-log theoretical calculation
arises from an uncertainty in the scale choice $Q$ at which 
effective theory decay matrix elements are evaluated. Varying $Q$
between $m_b\over 2$ to $2m_b$ leads to a theoretical uncertainty 
of $\sim 25$\%. 

Recently, next-to-leading order QCD corrections 
have been completed for $b\to s\gamma$ decay. These include {\it i}) 
complete virtual corrections\cite{ghw} to the relevant 
operators $O_2,\ O_7$ and $O_8$ which, when combined with 
bremsstrahlung corrections\cite{brem,ghw}
results in cancellation of associated soft and collinear singularities;
{\it ii}) calculation of ${\cal O}(\alpha_s^2)$ contributions to 
the ADM elements $\gamma_{ij}^{(1)}$
for $i,j=1-6$ (by Ciuchini {\it et al.}\cite{ciuchini}), for
$i,j=7,8$ by Misiak and M\"unz\cite{misiak1}, and for $\gamma_{27}^{(1)}$
by Chetyrkin, Misiak and M\"unz\cite{misiak2}. In addition, if 
two significantly different masses
contribute to the loop amplitude, then there can already exist significant
corrections to the Wilson coefficients at scale $M_W$. 
In this case, the procedure is to
create a tower of effective theories with which 
to correctly implement the RG running
between the multiple scales involved in the problem. 
The relevant operator bases, Wilson coefficients and
RGE's are given by Cho and Grinstein\cite{cg} for the SM and by
Anlauf\cite{anlauf} for the MSSM. The latter analysis includes contributions 
from just the $tW$, $tH^-$ and $\tst_i\tw_j$ loops 
(which are the most important ones). We include the above set of QCD 
improvements (with the exception of $\gamma_{27}^{(1)}$, which has been shown 
to be small\cite{misiak2}) 
into our calculations of the $b\to s\gamma$ decay rate for the mSUGRA model.

The contributions to $C_7(M_W)$ and $C_8(M_W)$ from $\tg\tq$ and $\tz_i\tq$
loops (SUSY contributions to $C_2(M_W)$ are suppressed by additional
factors of $g_s^2$)
have been presented in Ref. \cite{masiero}, although some
defining conventions must be matched between Ref. \cite{masiero} 
and Ref. \cite{anlauf} and Ref. \cite{ghw}. The only complication
is that the squark mixing matrix $\Gamma$ which enters the couplings 
must be derived. To accomplish this, we incorporate the following procedure
into our program for renormalization group running.
\begin{itemize}
\item We first calculate the values of all running fermion masses in the SM
at the mass scale $M_Z$. From these, we derive the corresponding 
Yukawa couplings $h_u$, $h_d$ and $h_e$ for each generation, and construct 
the corresponding Yukawa matrices $(h_u)_{ij}$, $(h_d)_{ij}$ and $(h_e)_{ij}$,
where $i,j=1,2,3$ runs over the 3 generations. We choose a basis that
yields flavor
diagonal matrices for $(h_d)_{ij}$ and $(h_e)_{ij}$, whereas the CKM mixing
matrix creates a non-diagonal matrix $(h_u)_{ij}$\cite{diego}.
\item The three Yukawa matrices are evolved within the MSSM from $Q=M_Z$
up to $Q=M_{GUT}$ and the values are stored. We use 1-loop RGEs except
for the evolution of gauge couplings.
\item At $Q=M_{GUT}$, the matrices $(Ah_u)_{ij}$, $(Ah_d)_{ij}$ and
$(Ah_e)_{ij}$ are constructed (assuming $A(M_{GUT})=A_0 \times {\bf 1}$). The
squark and slepton mass squared matrices $(M^2_k)_{ij}$ are also
constructed, where
$k=\tilde Q, \tilde u, \tilde d, \tilde L $ and $\tilde e$. These  
matrices are assumed diagonal at $Q=M_{GUT}$ with entries $m^2_0\delta_{ij}$.
\item The $(Ah)_{ij}$ and $(M^2_k)_{ij}$ matrices are evolved along with
the rest of the gauge/Yukawa couplings and soft SUSY breaking terms
between $M_Z$ and $M_{GUT}$ iteratively via Runge-Kutta method 
until a stable solution is found. The entire solution requires
the simultaneous solution of 134 coupled RGE's (with some slight redundancy).
We use 1-loop RGEs except for the evolution of gauge couplings.
\item At $Q=M_Z$, the $6\times 6$ $d$-squark mass squared 
matrix is constructed.
Numerical diagonalization of this matrix yields the squark mass mixing matrix
$\Gamma$ which is needed for computation of the $\tg\tq$ and $\tz_i\tq$
loop contributions.
\end{itemize}
At this point, the Wilson coefficients $C_7(M_W)$ and $C_8(M_W)$ can
be calculated and evolved to $Q\sim m_b$ as described above, so that the
$b\to s\gamma$ decay rate can be calculated \cite{bsg1}.

As an example, we show in Fig. \ref{nfig1} the calculated contributions 
to the Wilson coefficient $C_7(M_W)$ versus $\tan\beta$ for the mSUGRA
point $m_0,m_{1/2}=100,200$ GeV, $A_0=0$ and $\mu >0$. In frame {\it a}),
we show contributions from $\tw_i\tq$ loops, as well as from $tW$ and $tH^-$.
The $tW$ contribution is of course constant, while the $tH^-$ 
contribution is of the same sign, and increasing slightly in magnitude.
The various contributions from chargino loops increase roughly linearly with
$\tan\beta$ at a much faster rate, and thus form the dominant 
components of the $b\to s\gamma$ decay amplitude. In the case shown, there
are several large negative as well as positive contributions, so that
significant cancellations take place. 
The sum of all chargino loop contributions is shown by the dotted curve. 
In frame {\it b}) we show the 
contributions to $C_7(M_W)$ from different $\tg\tq$ loops. These contributions
vary with $\tan\beta$ as well and are comparable to
corresponding contributions from chargino loops. 
The sum of all gluino 
loop contributions is shown by the dotted curve; in this case, however,
the cancellation amongst the various loop contributions is nearly complete.

In frame {\it c}), we show the individual and summed 
contributions from $\tz_i\tq$ loops. 
These also increase with $\tan\beta$ but, as expected, are tiny compared
to the $\tw_i\tq$ and $\tg\tq$ loop contributions. The sum is again shown 
by the dotted curve. Here, the cancellations are not as complete as in the 
gluino loop case due to the Higgsino interactions of the neutralinos
which increase with $\tan\beta$.

\section{Numerical results and Implications}

\subsection{Constraints from $b\to s\gamma$ decay}

Our first numerical results for $B(b\to s\gamma )$ decay are shown in 
Fig. \ref{nfig2}, where we plot the branching fraction versus $\tan\beta$
for the mSUGRA point $m_0,m_{1/2}=100,200$ GeV, $A_0=0$ and for {\it a}) 
$\mu <0$ and {\it b}) $\mu >0$. The SM value, after QCD corrections, is
$B(b\to s\gamma )=3.2\pm 0.3\times 10^{-4}$, where the error comes from
varying the scale choice ${m_b\over 2}<Q<2m_b$\cite{bsg1,bsgsm}. 
The SM result is denoted by the dot-dashed line, and of course does not vary
with $\tan\beta$. If we include in addition the contribution from the
$tH^-$ loop, then we obtain the dotted curves, which always increase 
the value of $B(b\to s\gamma )$. For this parameter space point,
including the $tH^-$ loop always places the value of $B(b\to s\gamma )$
above the CLEO 95\% CL excluded region of $B(b\to s\gamma )<4.2\times 10^{-4}$.

If we include the full contribution of supersymmetric particles to the
computation of $B(b\to s\gamma )$, then we arrive at the solid curves in 
Fig. \ref{nfig2}. For the $\mu <0$ case in frame {\it a}), the SUSY loops
increase the branching fraction, which increases with $\tan\beta$,
so that the CLEO restriction on $B(b\to s\gamma )$ severely constrains
the mSUGRA model for large $\tan\beta$. For the frame {\it b}) case
with $\mu >0$, the SUSY loop contributions generally act to decrease 
the branching fraction, so that much of the parameter space is allowed
for moderate values of $\tan\beta$. Ultimately, as $\tan\beta$ increases, 
the decrease in $B(b\to s\gamma )$ becomes so severe that the mSUGRA model
becomes in conflict with the CLEO lower 95\% CL bound that
$B(b\to s\gamma )>1\times 10^{-4}$, so that for this particular mSUGRA
point, all values of $\tan\beta >21$ are excluded for the particular
choice of mSUGRA parameters.

In Figure \ref{nfig3}, we show the main result of this paper: the 
contours of constant $B(b\to s\gamma )$ in the $m_0\ vs.\ m_{1/2}$
parameter plane for large $\tan\beta =35$, for $A_0=0$ and for
{\it a}) $\mu <0$ and {\it b}) $\mu >0$. 
The contours are evaluated at a renormalization scale choice
$Q=m_b$.
The region marked by TH
is disallowed by theoretical considerations: either electroweak symmetry is 
not properly broken (the large $m_0$, small $m_{1/2}$ region) or
the lightest neutralino $\tz_1$ is not the lightest SUSY particle (LSP).
For small $m_0$, the light tau slepton $\ttau_1$becomes so light that
in the TH region, $m_{\ttau_1}< m_{\tz_1}$. The region denoted by EX
is excluded
by LEP2 constraints which require that 
the light chargino mass $m_{\tw_1}>80$ GeV.

In frame {\it a}), we see the value of $B(b\to s\gamma )$ is large
throughout the entire parameter space plane. The region with small 
values of $m_0$ and $m_{1/2}$ which is most favored by fine-tuning 
considerations\cite{ac} is in the most severe violation of the CLEO
constraint. The region below the dotted contour is in violation of the
CLEO 95\% CL bound for {\it all} choices of renormalization scale
${m_b\over 2}<Q<2m_b$. Thus, SUSY models allowed by CLEO for $\tan\beta =35$
and $\mu <0$ would be required to have $m_{\tg}>1470$ GeV 
(at $m_0,m_{1/2}=1000,600$ GeV)
and $m_{\tq}>1600$ GeV (for $m_0,m_{1/2}=320,800$ GeV).

In frame {\it b}) for $\mu >0$, we see that the values of 
$B(b\to s\gamma )$ are uniformly {\it below} the SM value, and so usually
in better agreement with the CLEO measured value of 
$B(b\to s\gamma )=2.32\pm 0.67\times 10^{-4}$ (where errors have been 
combined in 
quadrature). 
In fact, we note that the region with $m_{1/2}\simeq 500$ GeV agrees 
with the CLEO central value for $B(b\to s\gamma )$! This
region corresponds to parameter space points with $m_{\tg}\simeq 1200$ GeV,
and $m_{\tw_1}\simeq 400$ GeV.
In this frame, the entire plane shown, except the region below
the dotted contour, is {\it allowed} by the CLEO constraint. The region
below the dotted contour falls below the CLEO 95\% CL value of
$B(b\to s\gamma )>1\times 10^{-4}$ for all values of scale choice
${m_b\over 2}<Q<2m_b$. This is again the region most favored by 
fine-tuning. In this plane, $m_{\tg}\agt 525$ GeV and $m_{\tq}\agt 575$ GeV.

Up to this point, we have only shown results for a constant value
of $A_0=0$. In Figure \ref{nfig4}, we show contours of constant
$B(b\to s\gamma )$ in the $m_0\ vs.\ A_0$ plane for $m_{1/2}=200$ GeV,
for $\tan\beta =35$ and {\it a}) $\mu <0$ and {\it b}) $\mu >0$.
For frame {\it a}), we see that the branching fraction can change
by typically a factor of 2 over the parameter range shown, with most
of the variation occuring for changes in $m_0$, instead of with $A_0$.
The entire plane shown in frame {\it a}) is excluded by the CLEO bound.
In frame {\it b}), for $\mu >0$, the branching fraction can change by 
up to a factor of $\sim 4$ over the plane shown, again with most of
the variation coming due to changes in $m_0$.
In this case, the region to
the left of the dotted contour is excluded by the CLEO bound for
all choices of renormalization parameter ${m_b\over 2} < Q < 2m_b$.

\subsection{Comparison with relic density and direct 
detection rates for neutralino dark matter}

An important constraint on the mSUGRA model comes from implications
for the relic density of dark matter in the universe. The idea here is 
that in the very early universe, the LSP (the lightest neutralino) 
was a constituent of the matter
and radiation assumed to be in thermal equilibrium at some very 
high temperature. As the universe expanded and cooled, the LSP's could
no longer be produced, although they could still annihilate with one
another. Upon further expansion, the neutralino flux
dropped to such low levels that further annihilations would rarely occur, and
a relic abundance of neutralinos was locked in.
These relic LSP's could make up the bulk of dark matter in the universe today.

The neutralino relic density is calculable as a function of mSUGRA model 
parameter space\cite{relic}. 
The relic density is usually parametrized in terms of $\Omega h^2$, where
$\Omega =\rho/\rho_c$, $\rho$ is the relic density, 
$\rho_c$ is the critical closure density of the universe 
($\rho_c={3H_0^2\over 8\pi G_N}$) and $H=100h$ km/sec/Mpc is the scaled Hubble
constant with $0.5\alt h\alt 0.8$.
A value of $\Omega h^2>1$ implies a universe with age less than 
10 billion years, in conflict with the ages of the oldest stars. 
If $\Omega h^2 <0.025$, then the relic density of neutralinos cannot 
even account for the dark matter required by galactic rotation curves.
Some popular cosmological models that account for the COBE cosmic
microwave background measurements as well as structure formation in the 
universe actually prefer a mixed dark 
matter (MDM) universe\cite{primack}, with a matter 
density ratio of 0.3/0.6/0.1 for a 
hot dark matter/cold dark matter/baryonic matter mix. In this case, values
of $\Omega h^2\simeq 0.15-0.4$ are preferred.

In Fig. \ref{nfig5}, we show contours of constant relic density $\Omega h^2$.
The region to the right of the solid $\Omega h^2=1$ contour is excluded by
the age of the universe constraint, while the region below the dotted
contour has $\Omega h^2 <0.025$. The region betwen the dashed-dotted contours
is favored by a MDM universe. The region excluded by 
CLEO data is below the solid contour labelled $b\to s\gamma$. 
In frame {\it a}), we see that combining the two constraints allows only a 
small patch of allowed parameter space with $m_{1/2}>700$ GeV and 
$300\alt m_0\alt 600$ GeV. Over almost all of this region, the relic density
$\Omega h^2>0.4$. In frame {\it b}), the $b\to s\gamma$ excluded region
hardly intersects with the MDM region, so that large regions of
parameter space are favorable for cosmology as well as for CLEO
constraints!

In Fig. \ref{nfig5} we also plot one contour for expected rates for
direct detection of neutralino dark matter via cryogenic dark matter
detectors\cite{bbdm}. The calculations have been performed for neutralino
scattering from a $^{73}Ge$ detector. Current experiments are sensitive
to detection rates of 1-10 events/day/kg of detector. The goal
of such experiments is to achieve a sensitivity of $\sim 0.01$ events/kg/day
by about the year 2000. Towards this end, we show the 0.01 event/kg/day
contour in the $m_0\ vs.\ m_{1/2}$ plane; below the contour the event rates
exceed the 0.01/kg/day benchmark. In frame {\it a}), we see that the 
region accessible to direct neutralino detection coincides with the region
with very large $B(b\to s\gamma )$ rates well beyond the CLEO 95\% CL limit.
However, in frame {\it b}), for $\mu >0$, there exists a significant region
with large direct detection rates, which is allowed by the CLEO 
constraint, and is also in the favorable cosmological region!

\subsection{Implications for collider experiments}

The LEP2 $e^+e^-$ collider is expected to reach a peak CM energy of
$\simeq 195$ GeV, which should allow SM Higgs bosons of mass
$m_{H_{SM}}\alt 95$ GeV to be explored. For the $\tan\beta =35$ value shown
in Fig. \ref{nfig5}, the light Higgs scalar $m_h\agt 110$ GeV over the
entire $m_0\ vs.\ m_{1/2}$ plane shown. Hence, for these values of $\tan\beta$,
we would expect no Higgs signals to be seen at LEP2. The reach of LEP2 via
$\ttau_1\bar{\ttau_1}$ and $\tw_1\overline{\tw_1}$ searches is shown by the
dashed contour just above the region marked EX. 
This contour is defined by requiring
$m_{\tw_1}=95$ GeV and $m_{\ttau_1}=85$ GeV. For both cases of $\mu <0$ and
$\mu >0$ shown in Fig. \ref{nfig5}, the LEP2 sparticle reach falls below 
both the $B(b\to s\gamma )$ excluded region, and below the $\Omega h^2=0.025$
contours. If we accept the mSUGRA model literally, then the prediction is
that LEP2 should see no evidence for either a Higgs or SUSY if 
$\tan\beta\simeq 35$.

The Fermilab Tevatron $p\bar p$ collider is expected to operate at 
$\sqrt{s}=2$ TeV in Run 2, and to amass $\sim 2$ fb$^{-1}$ of integrated
luminosity by use of the Main Injector (MI). 
Ultimately, experiments hope to acquire $\sim 25$fb$^{-1}$
of integrated luminosity under the TeV33 program.
Recently completed calculations of the reach of the Tevatron MI
for mSUGRA at large $\tan\beta\simeq 35$ show a maximal reach in
$m_{1/2}$ to $\simeq 150$ GeV in the $\etmiss +$jets channel\cite{bcdpt2}. 
A similar reach has been calculated for TeV33, and finds points with 
$m_{1/2}\simeq 175$ accessible. Comparing these regions
to Fig. \ref{nfig5} shows that, like LEP2, the reach of Tevatron MI and TeV33
are below both the $B(b\to s\gamma )$ excluded contour and the 
$\Omega h^2=0.025$ contour, making discovery of SUSY particles highly unlikely
for mSUGRA if $\tan\beta$ is large. Over much of the parameter space plane
in frame {\it b}) of Fig. \ref{nfig5}, however, $m_h\alt 120$ GeV, which
(optimistically) corresponds to the maximal reach for $h$ at TeV33. 
Hence, if mSUGRA is correct
and $\tan\beta \simeq 35$, then TeV33 experiments may see a hint of the 
Higgs boson in their data sample. 
Of course, the entire large $\tan\beta$ parameter space shown should be easily
visible in at least the jets$+\eslt$ channel at the CERN LHC, even
with modest integrated luminosity.

\section{Model dependence of $\tg\tq$ loop contributions}

It is well known that within the mSUGRA framework
the chargino loop gives the dominant SUSY
contribution to the amplitude for the decay $b \to s\gamma$.
This can also be seen from Fig.~\ref{nfig1} where we see that while the
contributions from the gluino loops are individually comparable (or even
larger!) than those from chargino loops, these cancel out almost
completely leaving only a small residual contribution. In contrast,
while there is indeed considerable cancellation amongst the various
chargino contributions, there is nonetheless a sizeable residue that
remains. We may understand the large cancellations among the gluino
contributions in analogy with the familiar GIM cancellation in the SM:
indeed such a cancellation would be exact if squarks were precisely
degenerate ({\it i.e.}, the squark mass matrix is proportional to the unit
matrix) because we can then, by a unitary transformation, align the
squark and quark mass matrices, so that the gluino-squark-quark vertex
is exactly flavour diagonal. Within the mSUGRA framework with universal
soft breaking squark mass matrices at the unification scale, squarks are
indeed (approximately) degenerate, and gluino loop contributions to the
flavour violating $b \to s\gamma$ decay are suppressed. 
The GIM-like cancellation that we have described above does not occur
when Yukawa couplings enter the calculation as occurs, for instance, via
the higgsino components of chargino and neutralino loops. In this case,
as can be seen from Fig.~\ref{nfig1},
the cancellation is indeed incomplete (particularly for the chargino
where the large top quark Yukawa coupling enters).

These considerations lead us to examine whether the breaking of the
degeneracy of the soft SUSY-breaking squark masses at the unification
scale so strongly upsets the delicate cancellations that it results in large
gluino contributions to the amplitude for $b \to s\gamma$ decay. Of
course, by allowing soft-breaking mass squared matrices with arbitrary
off-diagonal entries, it should be possible to get very large flavour
violating gluino interactions. The issue that we address, however is
whether large gluino contributions are possible
even if we choose these soft squark matrices to be diagonal at the
unification scale. As
we will soon see, the physics of this ansatz is basis-dependent.

To parametrize the breaking of the squark degeneracy we begin by noting
that we may always choose a quark basis so that {\it either} the down or
the up type Yukawa couplings are diagonal at the weak scale. We will
call these the $d$- and $u$- cases, respectively. Next, these couplings are
evolved to $M_{GUT}$, where both up and down Yukawa matrices have
off-diagonal components. In the $d$-case, the down type Yukawa matrices
get off-diagonal 
contributions just from the RGE, while the up type Yukawas start off
off-diagonal right at the weak scale; in the $u$-case, the situation is
reversed. Up to now, the choice to work in the $u$ or $d$ cases is
purely a matter of convention, and indeed in previous sections we have
used the $d$ case. This is, however, no longer the case if we
further assume that the soft breaking squark mass squared matrix is diagonal 
(but not a multiple of the identity) at the GUT scale. This is because the
transformation that takes us from the $d$-case to the $u$-case does not
leave the squark mass squared matrix diagonal (except in the case when
this matrix is $m_0^2\times \bf{1}$). The $u$- and $d$- cases are thus
physically distinct. 

We have, therefore, studied these two cases
separately. To keep things simple, we split only the $b$-squarks ($t_L$
splits with $b_L$, of course) keeping the others degenerate at
$m_0$. The splitting is given by a single parameter 
\begin{displaymath}
x=(m_{\tb}/m_0)^2,
\end{displaymath}
where $m_b$ and $m_0$ are soft squark masses at the GUT scale. Thus $x=1$
corresponds to the mSUGRA case. We further consider
three possibilities where ({\it i})~just $\tb_L$, ({\it ii})~just 
$\tb_R$ and ({\it iii})~both $\tb_L$
and $\tb_R$ masses are split from those of other squarks. 

The results of our calculation of the gluino contribution to
$C_7(M_W)$ where the squark degeneracy
is broken as described above is shown in Fig.~\ref{nfig6} for
the $d$-case labelled $d$-diagonal, and for the $u$-case,
labelled $u$-diagonal. In each frame, we have three curves labelled $L$,
$R$ and $LR$ for the cases where just left, just right, and both left
and right sbottom soft masses are different from $m_0$. 
In our calculation, we have chosen $m_0=500$~GeV, $m_{1/2}=200$~GeV, 
$\tan\beta = 35$, $A_0=0$ and $\mu>0$. We choose a large value of $m_0$
so that the squark masses are not dominated by $m_{\tg}$ (in which case
splitting due to non-universal soft mass term would be unimportant). 
Also, for the large value of
$\tan\beta$ the bottom Yukawa coupling is significant.
The following
features are worth noting.
\begin{itemize}
\item For $x=1$ the value of $C_7^{gluino}(M_W)$ is the same for the up and
down cases for reasons that we have already explained.

\item For non-degenerate squarks, gluino loop effects are significantly
larger in the $u$-case. This may be understood if we recall that in the 
$d$-case, the mixing of down type Yukawas at $M_{GUT}$
arises {\it only} due to RGE.

\item Non-degeneracy effects when just the right squarks are split show
only small variation with the non-degeneracy parameter $x$ because
flavour mixing in the right squark sector is suppressed. 
\item Somewhat surprising is the fact that despite the large degree of
non-degeneracy, the gluino contribution to $C_7(M_W)$, which increases
by up to an order of magnitude relative to that in the mSUGRA case,
never becomes really large. For our choice of parameters, this may be
partly due to the fact that gluinos and squarks are significantly
heavier than $W$ and $t$. This is not the complete reason though. In our
computation we find that the three main contributions (from the two
sbottoms and the $\ts_L$) cancel one another leaving a remainder that is
typically smaller than 10-15\% of the largest contribution. While this
cancellation is much less complete than in the mSUGRA case, we are
unable to give a simple argument for why such a cancellation occurs.
\end{itemize}

Finally, we comment on the neutralino contributions for the case of
non-degenerate squarks. We have already noted that cancellations among
various chargino contributions are incomplete because of the effect of
large higgsino couplings to the $t\tt$ system. For large values of
$\tan\beta$, we may expect a similar effect for neutralinos. We may
further guess that this effect is largest in the $u$-case where flavor
mixing does not originate solely in the RGE. We have checked that for
values of parameters in Fig.~\ref{nfig6} above
a small value of $x =0.0625$, the neutralino contribution to $C_7(M_W)$
is indeed enhanced by a factor of 6-7 above its mSUGRA value, and
further, that this enhancement is largely due to incompleteness in the
cancellation between various contributions. For very large values of
$x$, $C_7(M_W)$ is about 1.5 times its mSUGRA value, but opposite in sign.
We thus conclude that while the neutralino contribution is somewhat
sensitive to the splitting of squark masses, it never appears to become
very dominant.

To summarize the results of this Section, we see that with our
assumptions, gluino contributions to the amplitude for the $b \to s
\gamma$ decay never dominate the SUSY contribution. This contribution
may nonetheless be non-negligible even if gluinos and squarks are well
beyond the reach of the Tevatron (and its proposed upgrades) as seen in
Fig.~\ref{nfig6}. We emphasize though that our conclusion is special to
models where all the flavour violation in the gluino-squark-quark vertex
at the GUT scale comes from non-diagonal Yukawa interactions. Larger
contributions from gluino loops may be possible in other models. 

{\it Note added:} After completion of this manuscript, a related paper
by Blazek and Raby appeared on the topic of $b\to s\gamma$ constraints 
on $SO(10)$ SUSY models at large $\tan\beta$\cite{raby}. 
Since Ref. \cite{raby} adopts a particular $SO(10)$ framework and
does not include the radiative electroweak
symmetry breaking constraint, comparison of results between the two papers
is not straightforward. Also, we have not included $\tt_L-\tc_L$ mixing
included in Ref.~\cite{raby} in our evaluation of the chargino loop.


\acknowledgments

We are grateful to M. Drees and T. ter Veldhuis for helpful
conversations and comments.
We thank the Aspen Center for Physics for hospitality while a portion of this 
work was completed.
This research was supported in part by the U.~S. Department of Energy
under grant number DE-FG-05-87ER40319.

%

%
\newpage
%
%
\begin{figure}
\caption[]{
We plot the value of the Wilson coefficient $C_7(M_W)$ versus $tan\beta$,
where $m_0,m_{1/2}=100,200$ GeV, $A_0=0$ and $\mu >0$. 
In {\it a}), we show the contribution from the $tW$ and $tH^\pm$ loops,
as well as contributions from various loops containing charginos.
The $\tq$ contribution includes both $\tc_L$ and $\tu_L$ squarks.
In {\it b}), we show the corresponding contributions from loops 
containing gluinos, and in {\it c}) we show the contributions from loops
including neutralinos.
}
\label{nfig1}
\end{figure}
\begin{figure}
\caption[]{
We show the branching ratio $B(b\to s\gamma )$ versus $\tan\beta$ for the same 
parameter space point as in Fig. 1, but with {\it a}) $\mu <0$ and {\it b})
$\mu >0$. The curves shown are for Standard Model prediction (dot-dashed),
SM plus charged Higgs prediction (dots) and the complete SUSY 
calculation (solid).
}
\label{nfig2}
\end{figure}
\begin{figure}
\caption[]{
Plot of contours of constant branching ratio $B(b\to s\gamma )$ in 
the $m_0\ vs.\ m_{1/2}$
plane, where $\tan\beta =35$, $A_0=0$ and $m_t=175$ GeV. 
Each contour should be multiplied
by $10^{-4}$. Frame {\it a}) is for
$\mu <0$ and frame {\it b}) is for $\mu >0$.
The regions labelled by TH (EX) are excluded by theoretical (experimental)
considerations. The EX region corresponds to the LEP2 limit
of $m_{\tw_1}>80$ GeV for a gaugino-like chargino.
}
\label{nfig3}
\end{figure}
\begin{figure}
\caption[]{
Plot of contours of constant branching ratio $B(b\to s\gamma )$ in 
the $m_0\ vs.\ A_0$
plane, where $\tan\beta =35$, $m_{1/2}=200$ GeV and $m_t=175$ GeV. 
Each contour should be multiplied
by $10^{-4}$. Frame {\it a}) is for
$\mu <0$ and {\it b}) is for $\mu >0$.
}
\label{nfig4}
\end{figure}
\begin{figure}
\caption[]{
Plot of contours of constant neutralino relic density $\Omega h^2$ in 
the $m_0\ vs.\ m_{1/2}$
plane, where $\tan\beta =35$, $A_0=0$ and $m_t=175$ GeV. 
We also show the region excluded by CLEO data on $B(b\to s\gamma )$
searches (the region below the $b\to s\gamma$ solid contour). The region
accessible to direct neutralino dark matter detectors is below the 
$10^{-2}$ contours. The region accessible to LEP2 sparticle searches is
below the dashed contour in the lower-left.
}
\label{nfig5}
\end{figure}
\begin{figure}
\caption[]{
Contributions to $C_7(M_W)$ from the gluino- squark loop for the $d$-case
and $u$-case discussed in the text as a function of the non-degeneracy
parameter $x=\frac{m_{\tb}^2}{m_0^2}$. The curves labelled $L$, $R$ and
$LR$ refer to the cases where just $m_{\tb_L}$, $m_{\tb_R}$ and both
$m_{\tb_L,R}$ are different from the universal squark mass $m_0$ at the
unification scale. For reference, we remind the reader that $C_7^{SM}=-0.23$
}
\label{nfig6}
\end{figure}
\vfill\eject


\centerline{\epsfbox{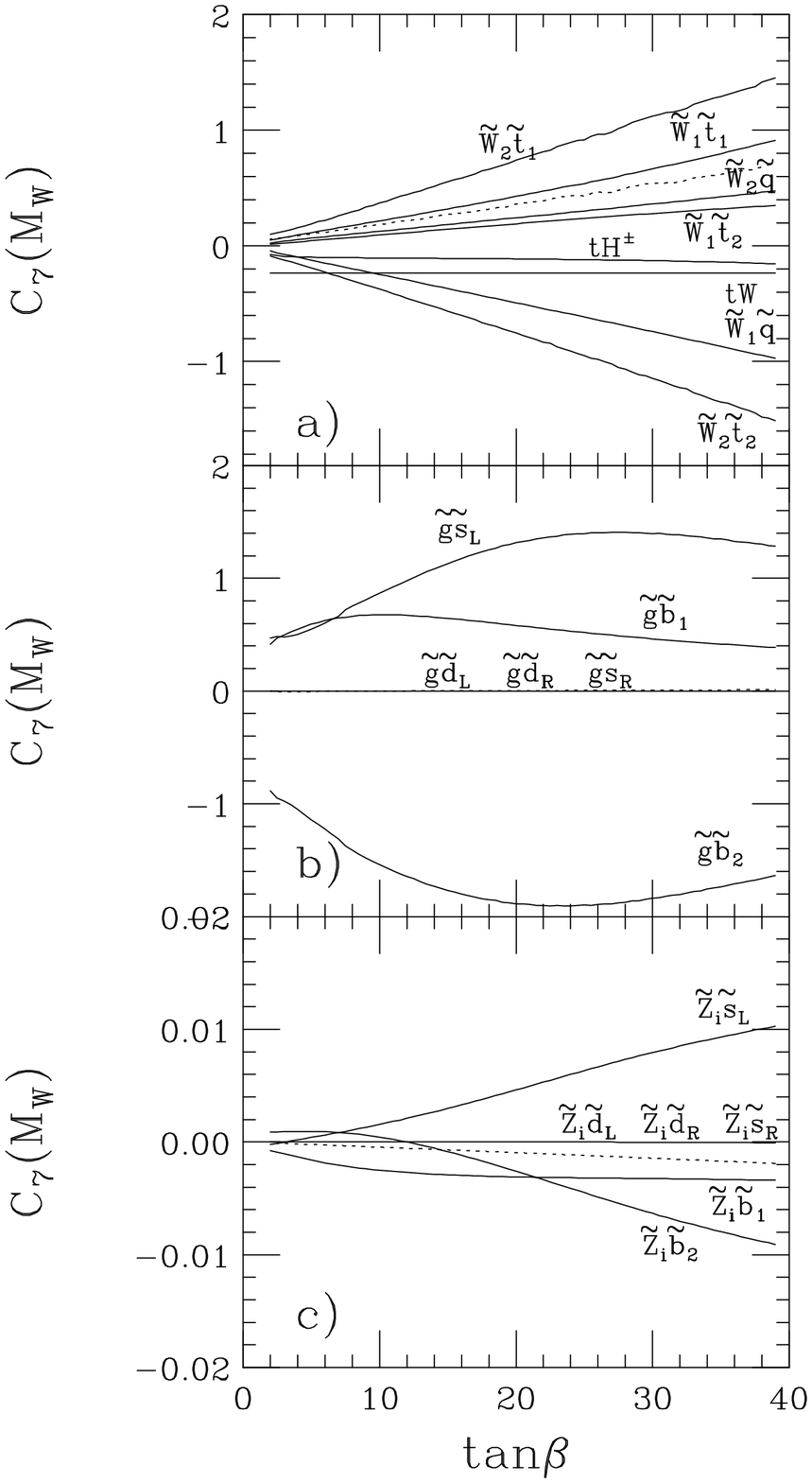}}
\vfill\eject

\centerline{\epsfbox{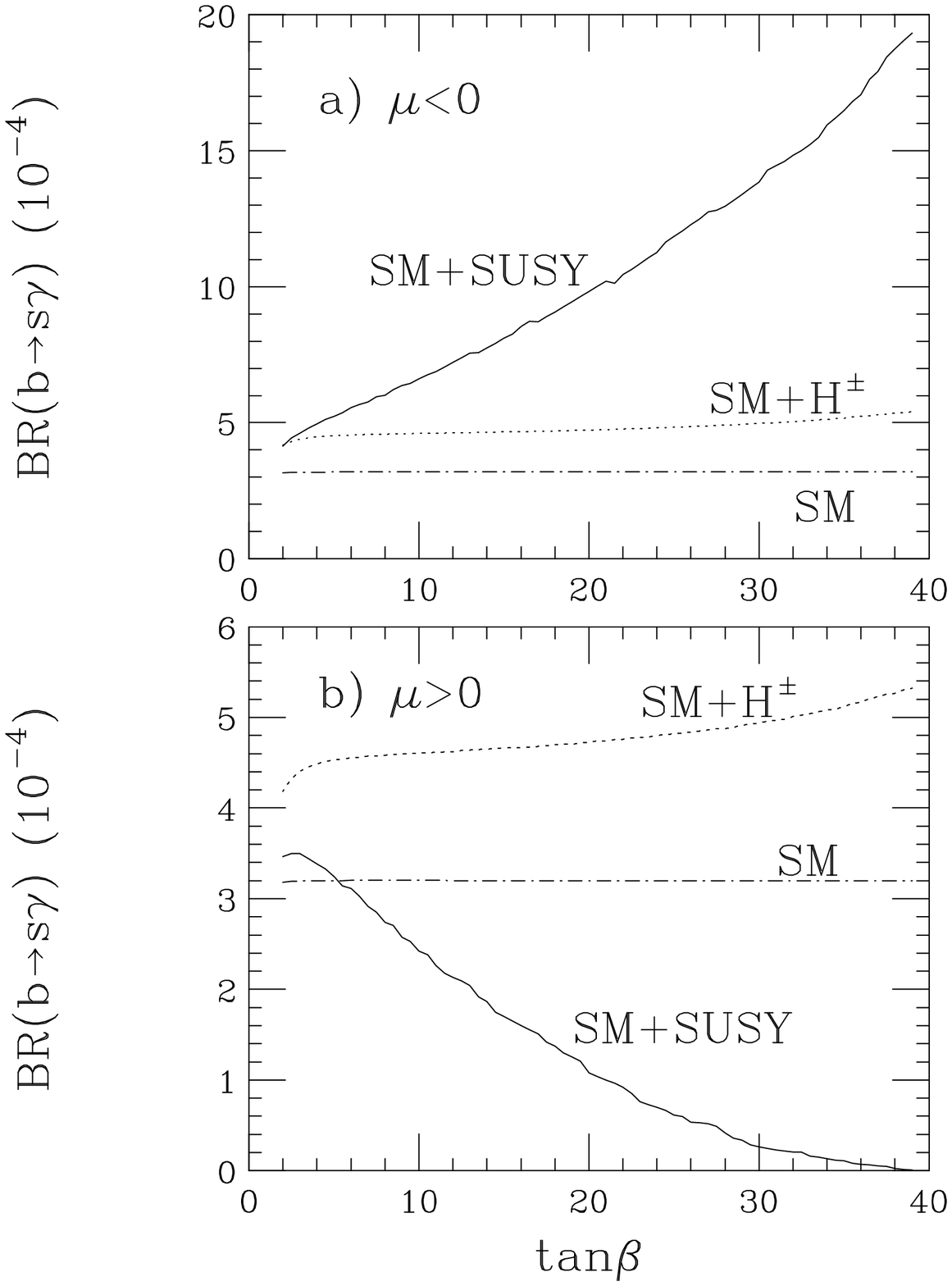}}
\bigskip\bigskip
\centerline{Fig.~2}
\vfill\eject

\centerline{\epsfbox{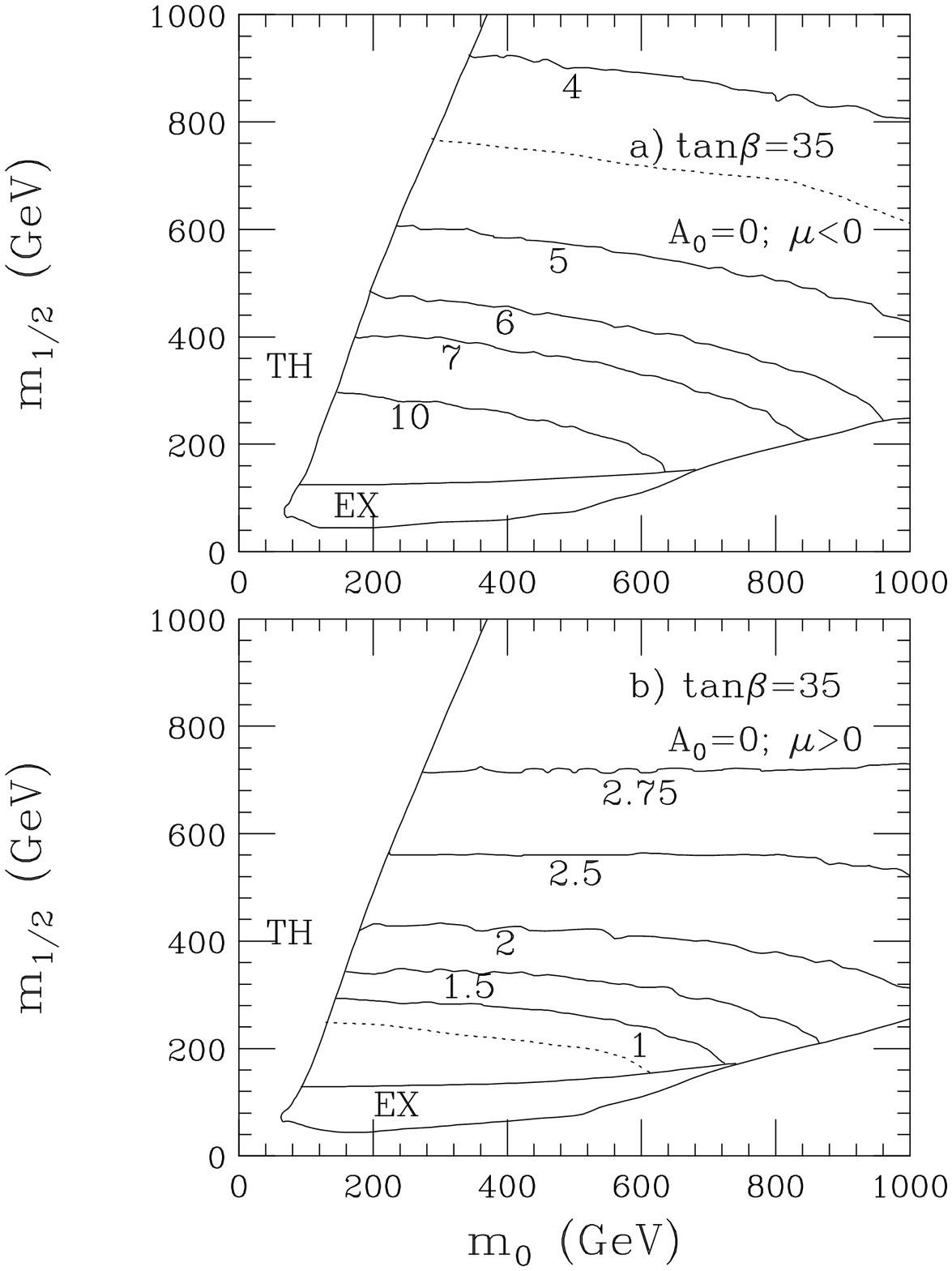}}
\bigskip\bigskip
\centerline{Fig.~3}
\vfill\eject

\centerline{\epsfbox{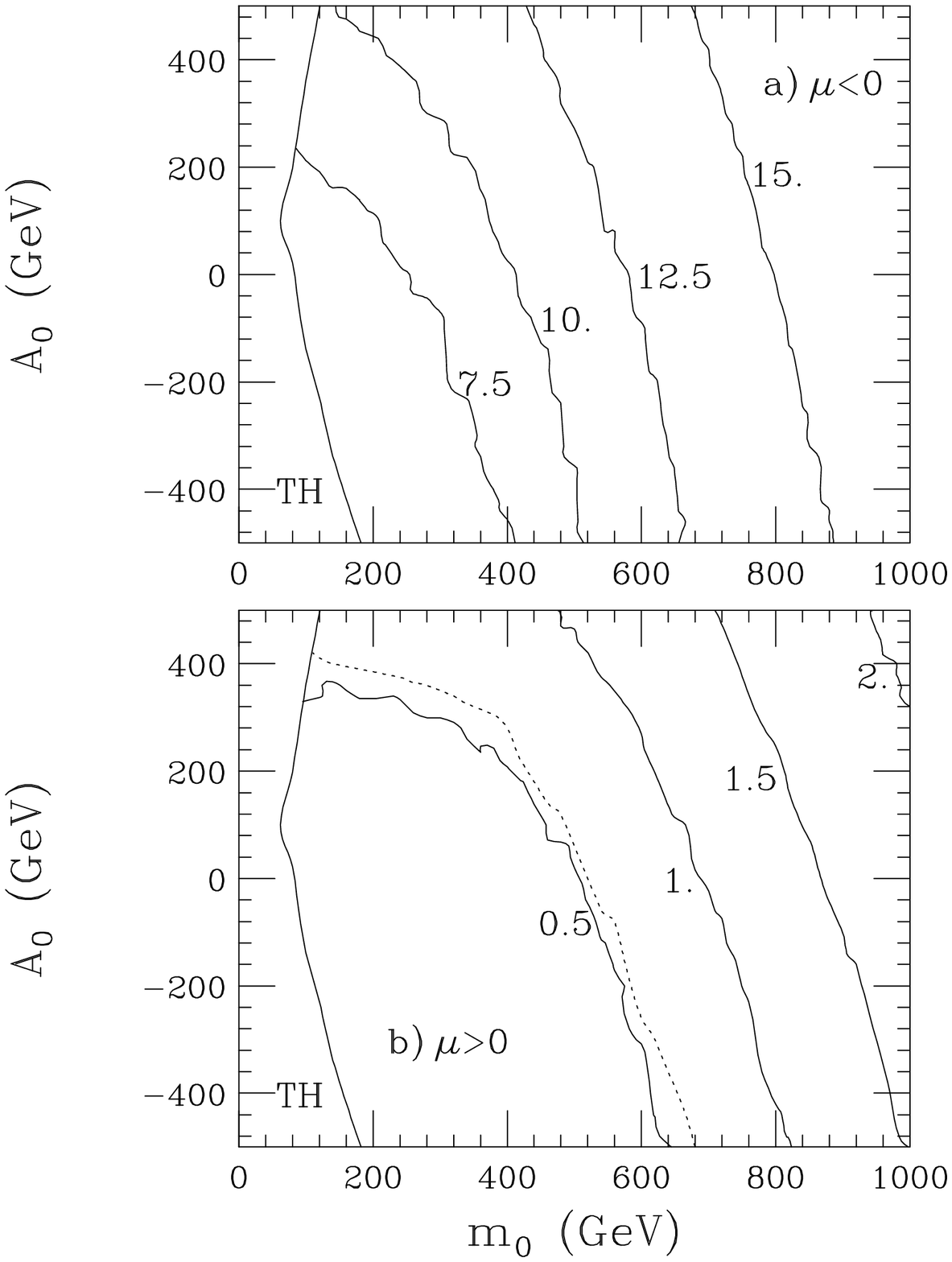}}
\bigskip\bigskip
\centerline{Fig.~4}
\vfill\eject

\centerline{\epsfbox{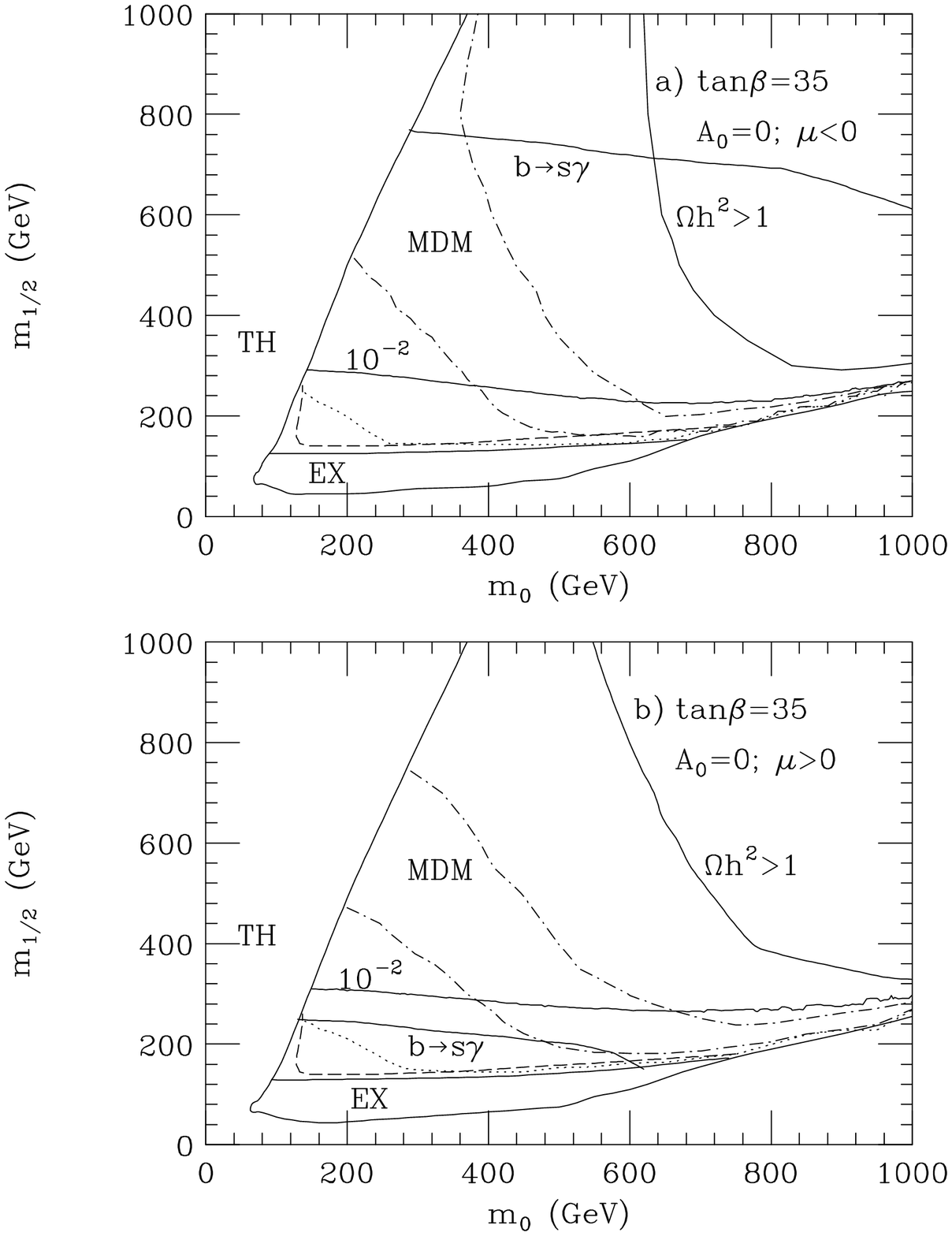}}
\bigskip\bigskip
\centerline{Fig.~5}
\vfill\eject

\centerline{\epsfbox{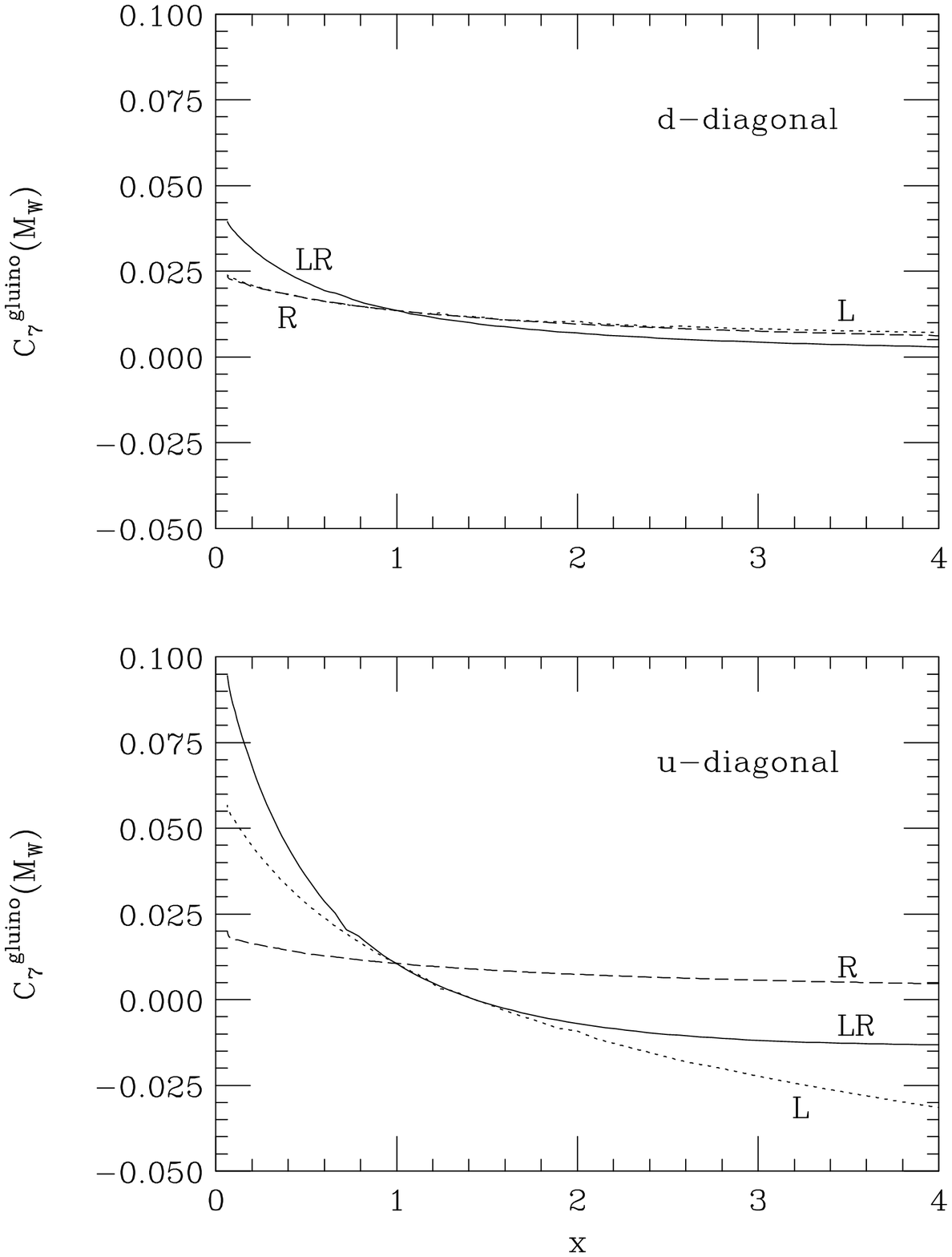}}
\bigskip\bigskip
\centerline{Fig.~6}
\vfill\eject


\begin{references}
%
\bibitem{haber}  See, {\it e.g.}, 
H. Haber in {\it Woodlands Superworld}, hep-ph/9308209 (1993). 
%
\bibitem{drees} See {\it e.g.}, M. Drees and S. Martin in 
{\it Electroweak Symmetry Breaking and New Physics at the TeV Scale}, 
edited by T. Barklow, S. Dawson, H~. Haber and J.~Seigrist, (World Scientific)
1995; see also J. Amundson {\it et al.} in {\it New Directions for High Energy 
Physics}, edited by D. G. Cassel, L. Trindle Gennari and R. H. Siemann,
(Stanford Linear Accelerator Center, 1996), hep-ph/9609374.
%
\bibitem{sugra} A. Chamseddine, R. Arnowitt and P. Nath, 
Phys. Rev. Lett. {\bf 49}, 970 (1982);
R. Barbieri, S. Ferrara and C. Savoy, Phys. Lett. {\bf B119}, 343 (1982);
L.J. Hall, J. Lykken and S. Weinberg, Phys. Rev. {\bf D27}, 2359 (1983).
%
\bibitem{cleo} M. S. Alam {\it et al.}, (CLEO Collaboration),
Phys. Rev. Lett. {\bf 74}, 2885 (1995). Note that recently 
the ALEPH collaboration
(at the International Europhysics Conference on High Energy Physics,
Jerusalem, Israel, Aug. 19-26, 1997) 
has made a preliminary announcement of a measurement of
$B(B\to X_s\gamma )=(3.38\pm 0.74\pm 0.85)\times 10^{-4}$. We do not
include this preliminary result in our analysis, and caution the reader
that some of our conclusions may change if the $b \to s \gamma$ decay
rate turns out to be closer to this ALEPH value.
%
\bibitem{masiero} S. Bertolini, F. Borzumati, A. Masiero and G. Ridolfi,
Nucl. Phys. B{\bf 353}, 591 (1991).
%
\bibitem{previous} R. Barbieri and G. F. Giudice, 
Phys. Lett. B{\bf 309}, 86 (1993); J. Lopez, D. Nanopoulos and G. Park, 
Phys. Rev. D{\bf 48}, 974 (1993); N. Oshimo, Nucl. Phys. B{\bf 404}, 20 (1993);
R. Garisto and J. Ng, Phys. Lett. B{\bf 315}, 372 (1993); 
M. Diaz, Phys. Lett. B{\bf 322}, 207 (1994);
Y. Okada, Phys. Lett. B{\bf 315}, 119 (1993);
F. Borzumati, Zeit. f\"ur Physik C{\bf 63}, 291 (1994);
P. Nath and R. Arnowitt, Phys. Lett. B{\bf 336}, 395 (1994);
G. Kane, C. Kolda, L. Roszkowski and J. Wells, 
Phys. Rev. D{\bf 49}, 6173 (1994);
F. Borzumati, M. Drees and M. Nojiri, Phys. Rev. D{\bf 51}, 341 (1995);
V. Barger, M. Berger, P. Ohmann and R. Phillips, 
Phys. Rev. D{\bf 51}, 2438 (1995);
F. Bertolini and F. Vissani, Zeit. f\"ur Physik C{\bf 67}, 513 (1995);
J. Lopez, D. Nanopoulos, X.~Wang and A.~Zichichi, 
Phys. Rev. D{\bf 51}, 147 (1995);
J. Wu, R. Arnowitt and P. Nath, Phys. Rev. D{\bf 51}, 
1371 (1995); 
B. de Carlos and J. A. Casas, Phys. Lett. B{\bf 349}, 300 (1995) 
and ERRATUM-{\it ibid} B{\bf 351}, 604 (1995).
\bibitem{bsg1} H. Baer and M. Brhlik, Phys. Rev. D{\bf 55}, 3201 (1997).
%
\bibitem{bcdpt} H. Baer, C. H. Chen, M. Drees, F. Paige and X. Tata,
Phys. Rev. Lett. {\bf 79}, 986 (1997).
%
\bibitem{carena}
M. Carena, J. Espinosa, M. Quiros and C. Wagner,
Phys. Lett. {\bf B355}, 209 (1995);
M. Carena, M. Quiros, C.E.M. Wagner,
Nucl. Phys. {\bf B461}, 407 (1996); H. Haber, R. Hempfling and A. Hoang,
Z. Phys. {\bf C}75, 539 (1997).
%
\bibitem{francesca} See F. Borzumati, Ref. \cite{previous}.
%
\bibitem{anlauf} H. Anlauf, Nucl. Phys. B{\bf 430}, 245 (1994).
%
\bibitem{gsw} See, {\it e.g.}, B. Grinstein, M. J. Savage and M. Wise,
Nucl. Phys. B{\bf 319}, 271 (1989); A. Ali, in {\it 20th International
Nathiagali Summer College on Physics and Contemporary Needs},
Bhurban, Pakistan, 1995, hep-ph/9606324.
%
\bibitem{ghw} C. Greub, T. Hurth and D. Wyler, 
Phys. Lett. B{\bf 380}, 385 (1996) and Phys. Rev. D{\bf 54}, 3350 (1996).
%
\bibitem{brem} A. Ali and C. Greub, Z. Phys. C{\bf 49}, 431 (1991),
Phys. Lett. B{\bf 259}, 182 (1991), {\bf 287}, 191 (1992) and 
{\bf 361}, 146 (1995);
Z. Phys. C{\bf 60}, 433 (1993); N. Pott, Phys. Rev. D{\bf 54}, 938 (1996);
%
\bibitem{ciuchini} M. Ciuchini, E. Franco, G. Martinelli and  L. Reina,
Nucl. Phys. B{\bf 415}, 403 (1994).
%
\bibitem{misiak1} M. Misiak and M. M\"unz, Phys. Lett. B{\bf 344}, 308 (1995).
%
\bibitem{misiak2} K. G. Chetyrkin, M. Misiak and M. M\"unz, 
Phys. Lett. {\bf B400}, 206 (1997).
%
\bibitem{cg} P. Cho and B. Grinstein, Nucl. Phys. B{\bf 365}, 279 (1991).
%
\bibitem{diego} H. Arason {\it et al.}, Phys. Rev. D{\bf 46}, 3945 (1992).
%
\bibitem{bsgsm} Ref. \cite{misiak2}, A. Buras, A. Kwiatkowski
and N. Pott, hep-ph/9707482 (1997) and C. Grueb and T. Hurth,
hep-ph9708214 (1997) quote a somewhat higher SM value of
$B(b\to s\gamma )$ than is presented here. The above works include
a 3\% increase due to non-perturbative effects calculated by
M. B. Voloshin, Phys. Lett. {\bf B}397, 275 (1997) plus the corrected
value of $\gamma_{27}$ in their calculations, which conspire to increase
the SM decay rate by about 10\%. These effects would similarly increase
our estimate of the SUSY contributions.
%
\bibitem{ac} G. Anderson and D. Casta\~no, Phys. Lett. {\bf B347}, 300 (1995)
and Phys. Rev. D{\bf 52}, 1693 (1995); see also 
K. L. Chan, U. Chattopadhyay and P. Nath,  hep-ph/9710473 (1997).  
%
\bibitem{relic} For a recent review, see
G. Jungman, M. Kamionkowski and K. Griest,
Phys. Rep. {\bf 267}, 195 (1996). See also M. Drees and M. Nojiri,
Phys. Rev. {\bf D47}, 376 (1993). The result shown here are from
H. Baer and M. Brhlik, Phys. Rev. {\bf D53}, 597 (1996). Further
references are included in these reports.
%
\bibitem{primack} See {\it e.g.}, J. Primack, astro-ph/9707285 (1997).
%
\bibitem{bbdm} H. Baer and M. Brhlik, Phys. Rev. D, in press,
hep-ph/9706509.
%
\bibitem{bcdpt2} H. Baer, C. H. Chen, M. Drees, F. Paige and X. Tata,
manuscript in preparation.
%
\bibitem{raby} T. Blazek and S. Raby, hep-ph/9712257 (1997).

\end{references}
\end{document}